\documentclass[letterpaper,english,twocolumn,aps,prl,superscriptaddress,10pt]{revtex4-1}
\usepackage[T1]{fontenc}
\usepackage[latin9]{inputenc}
\usepackage{booktabs}
\usepackage{amsmath}
\usepackage{amssymb}
\usepackage{graphicx}
\usepackage{color}

\makeatletter

\providecommand{\tabularnewline}{\\}

\makeatother

\usepackage{babel}
\begin{document}

\title{
Josephson current signatures of the Majorana flat bands on the surface of time-reversal-invariant Weyl and Dirac semimetals
}

\author{Anffany Chen}

\affiliation{Department of Physics and Astronomy, University of British Columbia, Vancouver, BC, Canada V6T 1Z1}
\affiliation{Quantum Matter Institute, University of British Columbia, Vancouver, BC, Canada V6T 1Z4}

\author{D.I. Pikulin}
\affiliation{Department of Physics and Astronomy, University of British Columbia, Vancouver, BC, Canada V6T 1Z1}
\affiliation{Quantum Matter Institute, University of British Columbia, Vancouver, BC, Canada V6T 1Z4}
\affiliation{Station Q, Microsoft Research, Santa Barbara, California 93106-6105, USA}

\author{M. Franz}
\affiliation{Department of Physics and Astronomy, University of British Columbia, Vancouver, BC, Canada V6T 1Z1}
\affiliation{Quantum Matter Institute, University of British Columbia, Vancouver, BC, Canada V6T 1Z4}

\date{\today}
\begin{abstract}
A linear Josephson junction mediated by the surface states
of a time-reversal-invariant Weyl or Dirac semimetal localizes Majorana flat
bands protected by the time-reversal symmetry. We show that as a result, the Josephson current exhibits a discontinuous jump at $\pi$ phase difference which can serve as an experimental signature of the Majorana bands. The magnitude
of the jump scales proportionally to the junction width and the momentum space distance between the Weyl nodes.  It also exhibits a
characteristic dependence on the junction orientation. We demonstrate
that the jump is robust against the effects of non-zero temperature
and weak non-magnetic disorder.

\end{abstract}
\maketitle

\section{\label{sec:level1}Introduction}
Curious particles that are their own antiparticles and potential
building blocks for quantum computing \cite{Sarma2015}, Majorana
zero modes (MZMs) have drawn intense interest and motivated quite a few proposals for realization in various topological or nodal superconductor systems \cite{Kitaev2001,Fu2008,Fu2009,lutchyn2010,oreg2010,Qi2011,Jason2012,Mourik2012,Tanaka2012,Schnyder2012,Meng2012,Cho2012,Wang2012,Beenakker2013,Wong2013,You2013,Li2014,Nadj-Perge2014,Hosur2014,Elliott2015,Ikegaya2015,Lu2015}.
For example, the setup obtained by proximitizing with a superconductor
the edge of a 2D topological insulator (TI) is predicted to host
MZMs at a Josephson junction implemented on the edge \cite{Fu2008,Fu2009}. 
When the phase difference across the junction is $\pi$, the lowest Andreev bound state (ABS) at the junction crosses zero energy and can be represented by two unpaired MZMs. For arbitrary phase difference the energy of the ABS is proportional to $\text{cos}(\varphi/2)$.
Since the ABS can
have positive or negative energy depending on $\varphi$, naively
thinking, the ABS would alternate between being occupied and unoccupied
as $\varphi$ varies in order to minimize the energy. Nonetheless,
because the ABS is nondegenerate,
whether it is occupied or not is determined by the fermion number
parity (an odd parity means it is occupied and vice versa) in the absence of quasiparticle poisoning. The many-body ground state of the system is thus $4\pi$-periodic because it has contributions from
this special $4\pi$-periodic ABS, on top of the usual $2\pi$-periodic ABSs at higher
energies and $\varphi$-independent quasiparticles above the gap
$\Delta_{0}$. The $4\pi$-periodicity is even more obvious in the limit of short junction, where only the
lowest ABS remains with energy given by 
\begin{equation}
E(\varphi)=\Delta_0 \text{cos}(\varphi/2)\label{eq:ABS_energy}
\end{equation}
where $\Delta_0$ is the superconducting gap.
As a consequence, the Josephson current exhibits
a characteristic $4\pi$-periodicity, in contrast to the $2\pi$-periodicity
for conventional Josephson junctions. This so-called fractional Josephson
effect would be a substantial evidence for MZMs \cite{Tanaka1997,kwon2004,Fu2009,Tanaka2009,Olund2012,Ilan2014}. So far the experiments have shown first signs of possible $4\pi$-periodicity \cite{Ilichev1998,rokhinson2012,Kurter2015,wiedenmann2016,Bocquillon2016}.

Different frontiers, Weyl and Dirac semimetals have also drawn
much attention lately as gapless topological phases beyond the usual classification of gapped TIs \cite{Turner2013,Ludwig2015}. Their
appeal is further stimulated by recent experimental observations of
their characteristic Fermi-arc surface states and negative magnetoresistance
due to chiral anomaly \cite{Liang2014,Huang2015,Lv2015,Xu2015,Xu2015a,Xu2015b,Zhang2016}.
In this work, we show that the fractional Josephson effect has implications
in a time-reversal-invariant Weyl semimetal (WSM) (time-reversal breaking case is discussed in \cite{Khanna2016,Baireuther2017}). For our purposes, a Dirac semimetal can be considered as a special
case of Weyl semimetal with coinciding Weyl nodes due to point group
symmetry. Hence the following argument on Weyl semimetals also holds
for Dirac semimetals. 

The argument goes as follows. The Fermi arcs can be regarded as helical
edge states of 2D TIs embedded in the $\mathcal{T}$-invariant WSM \cite{Turner2013}.
Proximitizing the surface of the $\mathcal{T}$-invariant WSM thus amounts
to proximitizing the edges of the 2D TIs. As demonstrated by the previous
work of two of the present authors \cite{Chen2016}, a short Josephson junction with $\pi$ phase difference on the surface of $\mathcal{T}$-invariant WSM would localize many pairs of decoupled MZMs
labeled by different momenta $k$, each pair due to an embedded 2D
TI. Naturally it seems that the fractional Josephson effect should
also carry over, but in this case it does not violate the fermion
parity to change the occupancies of ABSs. Indeed, a Cooper pair can
split and occupy ABSs at different $k$'s. The $4\pi$-periodicity
is no longer protected by parity; instead we have a distinct Josephson
current jump at odd-$\pi$ phase differences where many ABSs simultaneously
change their occupancies in order to minimize the ground state energy.

This phenomenon is present in any $\mathcal{T}$-invariant Weyl or Dirac
semimetal. Among all candidates, the Dirac semimetal Cd$_{3}$As$_{2}$
has the advantage of being experimentally accessible and having the
minimal number of four Weyl points \cite{Wang2013,Cano2016}. For concreteness
and experimental implication, in the following we focus our discussion
on this material. 

\begin{figure}
\centering{}\includegraphics[scale=0.4]{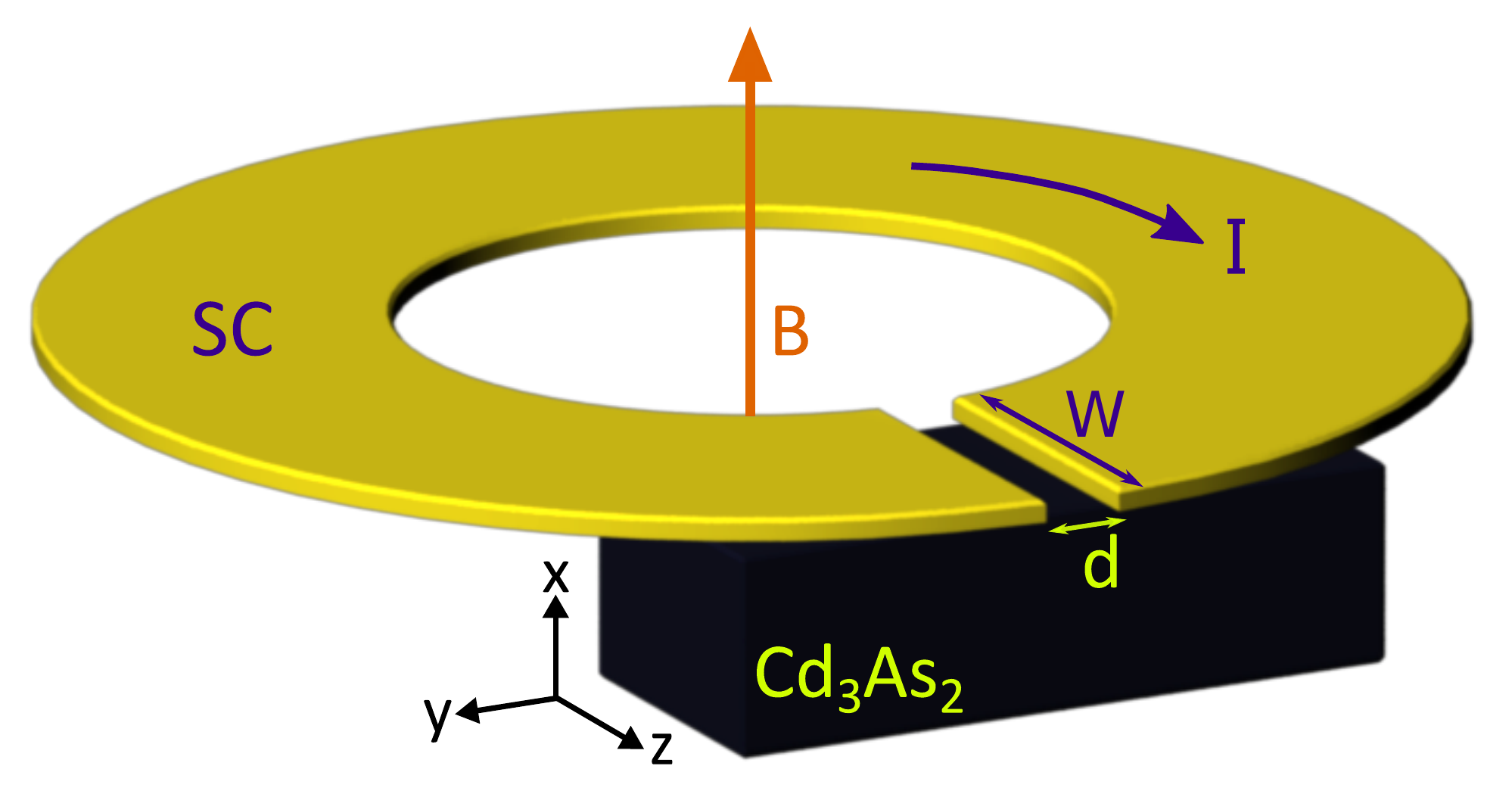}\caption{\label{fig:setup}By placing a superconducting ring with a narrow opening on top of a Cd$_3$As$_2$ slab, one obtains a Josephson junction mediated by the surface states of Cd$_3$As$_2$. We label the length of the junction by $d$ and the width by $W$. An external flux going through the ring generates a phase difference  $\varphi=2\pi(\Phi/\Phi_0)$ across the junction, which leads to supercurrent measurable by a capacitively-coupled SQUID sensor. To study the lowest ABS, we assume $d\ll \xi$, the proximity-induced coherence length.}
\end{figure}

Consider a linear Josephson junction obtained by proximitizing the top surface of a Cd$_3$As$_2$ slab with an open superconducting ring, as shown in Fig. \ref{fig:setup}. To study the lowest ABS, we work in the short-junction limit where the junction length $d\ll \xi$, the proximity-induced coherence length. An applied flux $\Phi$ through the ring creates phase difference $\varphi=2\pi(\Phi/\Phi_0)$ across the junction, where $\Phi_0=h/2e$ is the superconducting flux quantum. The phase difference drives a supercurrent around the ring, which can be detected by a SQUID sensor as in \cite{Sochnikov2013,Sochnikov2015}, or by including the junction into a SQUID with a stronger junction in parallel and measuring the modulation of the total supercurrent, as done for example in \cite{della2007}. 

We predict that the current-phase relation (CPR) exhibits a discontinuous
current jump at $\varphi=\pi$. The jump size is proportional to the
number of MZMs, which is maximized when the junction width is oriented
along the
direction in which the Dirac nodes are separated, and decreases as $\text{cos}(\theta)$
when the junction is rotated by an angle $\theta$. Low temperature and weak non-magnetic disorder smoothen the discontinuity to some extent, but the CPR profile remains highly skewed and the jump size is unaffected. We estimate
that a current jump on the order of 1$\mu$A can be observed in typical
experiments. 

Our prediction is supported by analysis in Sec. \ref{sec:level2}, where we demonstrate the existence of Majorana flat bands, compute and interpret the CPR, and investigate the effects of non-zero temperature and non-magnetic disorder.

\section{\label{sec:level2}Analysis and Results}

\subsection{\label{sec:level2-1}Majorana flat bands on the
surface of Cd$_{3}$As$_{2}$}

The low-energy effective Hamiltonian of tetragonal Cd$_{3}$As$_{2}$ has been extracted from symmetry considerations \citep{Wang2013,Cano2016}. In the basis $\{|P_{J=\frac{3}{2}},J_{z}=\frac{3}{2}\rangle$,
$|S_{\frac{1}{2}},\frac{1}{2}\rangle$, $|S_{\frac{1}{2}},-\frac{1}{2}\rangle$,
$|P_{\frac{3}{2}},-\frac{3}{2}\rangle\}$, where $S$ and $P$ refer
to spin-orbit-coupled Cd-5$s$ and As-4$p$ states, the $4\times4$
Hamiltonian takes the form 
\begin{equation}
H_{0}(\boldsymbol{k})=\epsilon_{0}(\boldsymbol{k})+\left(\begin{array}{cccc}
M(\boldsymbol{k}) & Ak_{-} & 0 & 0\\
Ak_{+} & -M(\boldsymbol{k}) & 0 & 0\\
0 & 0 & -M(\boldsymbol{k}) & -Ak_{-}\\
0 & 0 & -Ak_{+} & M(\boldsymbol{k})
\end{array}\right)\label{eq:H0}
\end{equation}
up to second order in $k$ near the $\Gamma$ point. $\epsilon_{0}(\boldsymbol{k})=C_{0}+C_{1}k_{z}^{2}+C_{2}(k_{x}^{2}+k_{y}^{2})$,
$M(\boldsymbol{k})=M_{0}+M_{1}k_{z}^{2}+M_{2}(k_{x}^{2}+k_{y}^{2})$,
$k_{\pm}=k_{x}\pm ik_{y}$. The parameters in the Hamiltonian are best-fit values
to the ab initio calculation \cite{Cano2016}, and are summarized in Table \ref{tab:parameters}.

\begin{table}
\begin{tabular}{|c|c|c|c|}
\hline 
$C_{0}$ {[}eV{]} & $C_{1}$ {[}eV$\textrm{\AA}^{2}${]} & $C_{2}$ {[}eV$\textrm{\AA}^{2}${]} & $A$ {[}eV$\textrm{\AA}${]}\tabularnewline
\hline 
-0.0145 & 10.59 & 11.5 & 0.889\tabularnewline
\hline 
\hline 
$M_{0}$ {[}eV{]} & $M_{1}$ {[}eV$\textrm{\AA}^{2}${]} & $M_{2}$ {[}eV$\textrm{\AA}^{2}${]} & $a$ {[}$\textrm{\AA}${]}\tabularnewline
\hline 
0.0205 & -18.77 & -13.5 & 20\tabularnewline
\hline 
\end{tabular}

\caption{\label{tab:parameters}Material parameters taken from \citep{Cano2016}
are used in our simulations. We let the effective lattice constants be $a_x=a_y=a_z=20\textrm{\AA}\equiv a$ rather
than the values $a_x=a_y=3\textrm{\AA}$ and $a_z=5\textrm{\AA}$ provided in  \citep{Wang2013}, so
that the relevant low-energy band structure is more numerically resolvable.}
\end{table}
 Each diagonal block describes a Weyl semimetal. Diagonalizing this
Hamiltonian gives the energy spectrum
\begin{equation}\label{eq3}
E(\boldsymbol{k})=\epsilon_{0}(\boldsymbol{k})\pm\sqrt{M(\boldsymbol{k})^{2}+A^{2}(k_{x}^{2}+k_{y}^{2})}
\end{equation}
The Weyl nodes are located at $(0,0,\pm Q)$ where $Q=\sqrt{-M_{0}/M_{1}}$ such that the square-root term vanishes. Each location has two Weyl
nodes of opposite chiralities, one from each diagonal block in Eq.
\eqref{eq:H0}, and the point group symmetry prevents them from mixing.
In our analysis, the chemical potential $\mu$ is always set to the
energy of Weyl nodes, $E_0=C_{0}-C_{1}M_{0}/M_{1}$ to minimize the influence of bulk states. 
\begin{figure*}\includegraphics[scale=0.3]{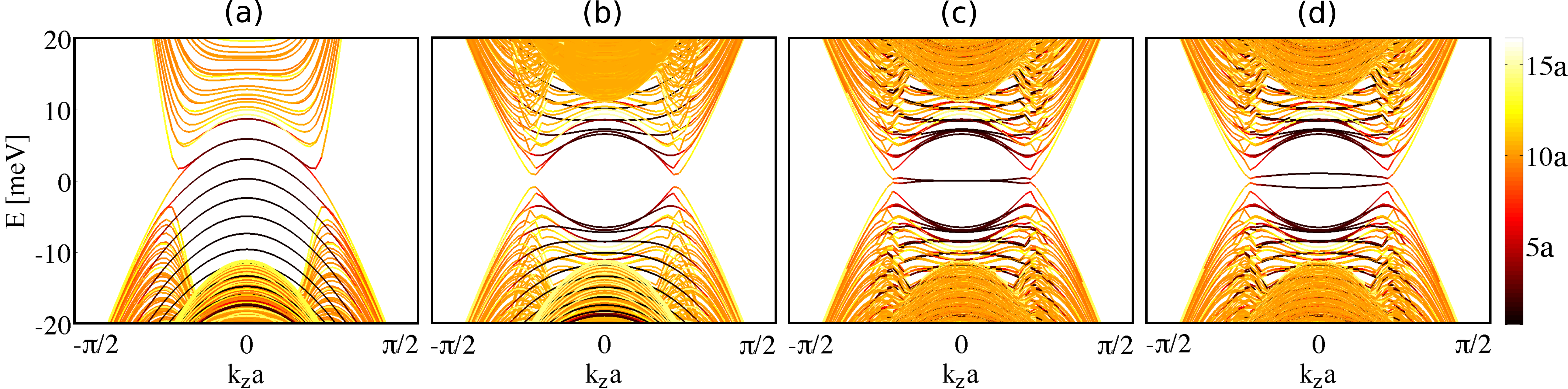}\caption{\label{fig:wide}Spectrum of the $L_{x}a\times L_{y}a\times L_{z}a=40a\times 52a\times 60a$ ($a=20\textrm{\AA}$) slab of Cd$_3$As$_2$ with periodic boundary condition in the $y$- and $z$-directions. The material parameters are listed in Table \ref{tab:parameters}. Energy states are color-coded according to the expectation value of distance from the surface. Panel (a) shows the normal-state energy spectrum with the Dirac cones separated along $k_{z}$ and the surface Fermi arcs spanned between them. In Panel (b) a uniform pairing potential of magnitude  $\Delta_{0}=10$meV is introduced on the surfaces; the Fermi arcs are gapped out. In Panel (c), two junctions of $\pi$ phase difference are implemented on the top
surface and the bottom surface remains uniformly superconducting;
the resulting zero-energy surface flat bands linking the Dirac nodes are four-fold
degenerate and decoupled. In Panel (d), the phase difference deviates
from $\pi$ to $1.1\pi$; the flat bands get split from zero energy and obtain small dispersion, while the rest of the bands stay the same.
In Panels (b)-(d), the small gaps near the Dirac nodes are due to the finite-size effect.}
\end{figure*}

To model a lattice, we turn $H_{0}(\boldsymbol{k})$
into a tight-binding Hamiltonian $H_{\text{tb}}(\boldsymbol{k})$ via the
substitutions $k_{i}^{2}\sim\frac{2}{a^{2}}(1-\mbox{cos}(k_{i}a))$
and $k_{i}\sim\frac{1}{a}\mbox{sin}(k_{i}a)$, where $a$ is the lattice
constant for the low-energy effective model. $Q$ now satisfies cos$(aQ)=1+(a^2/2)(M_0/M_1)$. We let $a=20\textrm{\AA}$ even though $a_x=a_y=3\textrm{\AA}$ and $a_z=5\textrm{\AA}$ are the best-fit values to the ab initio calculation \citep{Wang2013}. The larger lattice constants make the relevant low-energy band structure more numerical resolvable. This approach is not an issue because our numerics mainly serve illustrative purposes; later on the quantitative predictions will be made with realistic parameters. We consider a slab of Cd$_{3}$As$_{2}$ with $L_{x}\times L_{y}\times L_{z}$ sites. It is periodic in the $y$- and $z$-directions and finite in the
$x$-direction. We Fourier transform $H_{\text{tb}}(\boldsymbol{k})$ along the $x$-direction in order to define individual layers. Then
numerical diagonalization gives the normal-state energy spectrum,
showing two Dirac cones in the bulk and two Fermi arcs on each surface.
Fig. \ref{fig:wide}(a) projects the spectrum onto $k_{z}$-axis,
so the Fermi arcs overlap and appear as curved bands (each corresponding
to a different $k_{y}$) connecting the Dirac cones. 

Next we verify that the surface states can be gapped out by $s$-wave pairing of time-reversed states. We introduce a pairing potential to both top and bottom surfaces. The matrix of Bogoliubov-de
Gennes (BdG) Hamiltonian reads 
\begin{equation}
H_{\text{BdG}}=\left(\begin{array}{cc}
H_{\text{tb}}-\mu & \Delta\\
\Delta^{\dagger} & -TH_{\text{tb}}T^{-1}+\mu
\end{array}\right)\label{eq:BdG}
\end{equation}
where every entry is a $4L_{x}\times4L_{x}$ matrix whose basis are tensor products of spin-coupled orbitals and $x$-layers. $H_{\text{tb}}$
is as discussed above, and its time-reversal conjugate $TH_{\text{tb}}T^{-1}=H_{\text{tb}}$ because it is time-reversal invariant. The pairing potential $\Delta$ is a diagonal
matrix whose nonzero elements are $\Delta_{0}$ when the $x$-layer index is 1 or $L_{x}$. Diagonalizing $H_{\text{BdG}}$ gives a particle-hole symmetric
band structure with gapped-out surface Fermi arcs and intact bulk
Dirac cones, as shown by Fig. \ref{fig:wide}(b). 

Lastly we Fourier transform $H_{\text{tb}}$ again along $y$; now
each block in $H_{\text{BdG}}$ is a matrix whose basis are tensor
products of spin-coupled orbitals, $x$-layers, and $y$-sites. Then
we let the nontrivial diagonal elements of $\Delta$ be $\Delta_{0}$
when the $x$-layer index is $L_{x}$ and

\begin{equation}
\begin{cases}
\Delta_{0}e^{i\varphi} & \text{for }L_{y}/4<y/a\le3L_{y}/4\\
\Delta_{0} & \text{for \ensuremath{y/a\le L_{y}/4} or \ensuremath{y/a>3L_{y}/4}}
\end{cases}
\end{equation}
when the $x$-layer index is 1. This defines two evenly spaced short
junctions each of phase difference $\varphi$ on the top surface.
Two evenly spaced junctions are needed to satisfy the periodic boundary
condition along $y$. We keep the bottom surface uniformly superconducting
so that the bottom surface states do not obstruct the zero-energy
flat bands in the figures. Note that the junctions are periodic widthwise,
but a realistic junction would have a finite width. In principle, sides of the sample in Fig. 1 could also carry supercurrent, which is not accounted for in our model. However, in the geometry where the junction width is much larger than the thickness of Cd$_{3}$As$_{2}$ slab, the contribution of the side modes is negligible. This is additionally substantiated by Ref. \citep{Sochnikov2015}, a S/TI/S junction experiment
set up like Fig. 1: the authors did not consider this geometric effect, yet found
good agreement between their theoretical model and experimental data.
The energy spectrum of $H_{\text{BdG}}(\varphi=\pi)$ in Fig.
\ref{fig:wide}(c) clearly shows flat bands of zero-energy surface
states, which are Majorana or self-conjugate for reasons explained
in \citep{Chamon2010,Beenakker2015}. There are in total four Majorana
flat bands. Each junction locally hosts two such bands, which hybridize
and gain a non-zero energy when $\varphi$ moves away from $\pi$, as
shown by Fig. \ref{fig:wide}(d). As long as the time-reversal symmetry
$\mathcal{T}$ is respected, $\varphi$ must be either $0$ or $\pi$,
so we can say that the Majorana flat bands are protected by $\mathcal{T}$.
Moreover, translational symmetry prohibits MZMs in the same band from
mixing because $k_{z}$ is a good quantum number. Lastly, the flat
bands at different junctions are decoupled due to the artificial symmetry
of evenly spaced junctions.

The number of MZM pairs, $N$, in each junction is determined
by the junction width, $W=aL_{z}$, and the distance between the Dirac nodes,
$2Q$: 
\begin{equation}
N=WQ/\pi \label{eq:N}
\end{equation}
If the junction is rotated by an angle $\theta$,
\begin{equation}
N=WQ\text{cos}(\theta)/\pi.\label{eq:angle}
\end{equation}
because the distance between the Dirac nodes, as projected onto the 1D Brillouin zone parallel to the junction width, becomes $2Q\cos\theta$. In our numerical model, $aQ\sim0.21\pi$ and $L_z=60$, so $N\sim 13$ at each junction. In a realistic model of Cd$_{3}$As$_{2}$, $Q=\sqrt{-M_{0}/M_{1}}=0.033\textrm{\AA}^{-1}$. Assuming a typical junction width of $W=1\text{\ensuremath{\mu}m}$, one finds that $N\sim100$. Hence a significant MZM population can be easily obtained in experiments, especially with a wide junction.

\subsection{\label{sec:level2-2} Current-phase relation at non-zero temperature}

\begin{figure}
\includegraphics[scale=0.38]{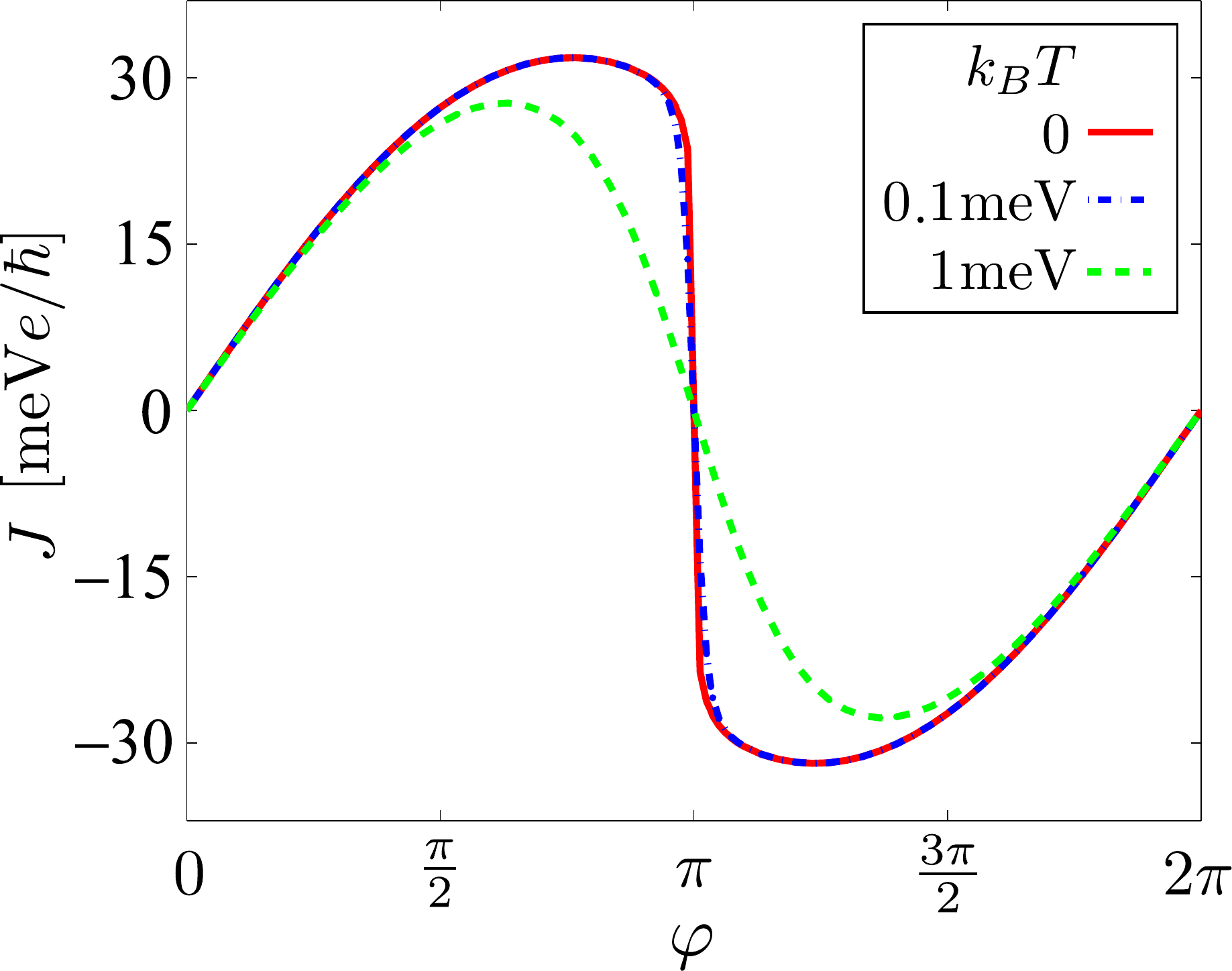}\caption{\label{fig:CPR}The Josephson current (per junction) computed from Eq. \eqref{eq:finite_temp} is shown at various $\varphi$. At $k_{B}T=0$
and 0.1meV, there is a clear current jump of magnitude $\sim60$meV$e/\hbar$
or $15\mu\text{A}$. At $k_{B}T=1$meV, the CPR retains little skewness. The magnitude of pairing potential is $\Delta_{0}=10$meV in the numerics, which is unrealistically large so that the superconducting gap is larger than the finite-size gaps that are always present in finite systems. $\Delta_{0}=0.1$meV is used later in quantitative analysis.}
\end{figure}

One can detect the Majorana flat bands by measuring
the CPR. The MZMs cause a current
jump at $\pi$ phase difference. The jump size is proportional to $N$. To demonstrate this in Cd$_{3}$As$_{2}$, we
compute the Josephson current given by

\begin{equation}
J(\varphi)=\frac{2e}{\hbar}\frac{\text{d}\varepsilon(\varphi)}{\text{d}\varphi}\label{eq:supercurrent}
\end{equation}
where $\varepsilon(\varphi)$ is the many-body ground state energy
of the system \citep{Nazarov2009}. We diagonalize $H_{\text{BdG}}(\varphi)$ for every fixed $\varphi$. The positive energies,
labelled by $E_{i}(\varphi)$, are the excitation energies of Bogoliubov quasiparticles. Due to particle-hole symmetry, the negative energies of the filled states below $E_{F}=0$ are simply $-E_{i}(\varphi)$, so
\begin{equation}
\varepsilon(\varphi)=-\frac{1}{2}\sum_{i}E_{i}(\varphi)
\end{equation}
where the factor of $\frac{1}{2}$ compensates the particle-hole doubling in the BdG construction. As shown in Fig. \ref{fig:CPR}, the resulting $J(\varphi)$ displays a $2\pi$-periodic CPR with a current jump  of magnitude $\sim60$meV$e/\hbar$ or 
$15\mu\text{A}$.

At non-zero temperature,
quasiparticles below $E_{F}=0$ can be thermally excited to occupy
positive energy states. With the free energy $F(\varphi)$ in place
of $\varepsilon(\varphi)$, the Josephson current becomes \citep{Bardeen1969,Beenakker1991}

\begin{equation}
F=-k_{B}T\text{ln}Z=-k_{B}T \sum_{i} \ln \bigg(2\cosh \bigg(\frac{E_{i}(\varphi)}{2k_{B}T}\bigg)\bigg)
\end{equation}
\begin{equation}
J(\varphi)=\frac{2e}{\hbar}\frac{\text{d}F(\varphi)}{\text{d}\varphi}=-\frac{e}{\hbar}\sum_{i}\frac{\text{d}E_{i}(\varphi)}{\text{d}\varphi}\text{tanh}\bigg(\frac{E_{i}(\varphi)}{2k_{B}T}\bigg) \label{eq:finite_temp}
\end{equation}
which agrees with Eq. \eqref{eq:supercurrent} when $T\rightarrow0$ \footnote{Our expression for Josephson current differs from that in Refs. \citealp{Bardeen1969,Beenakker1991}
by a factor of 2, which was there to account for spin degeneracy.
In our case, any degeneracy is already included in the summation because
we assign a unique label to every eigenstate.}.
Fig. \ref{fig:CPR} shows that the current jump is slightly rounded
at $k_{B}T/\Delta_0=0.01$ but significantly loses skewness when $k_{B}T/\Delta_0=0.1$. In experiments, one should aim at lowering $k_{B}T/\Delta_0$ to 1$\%$. For a typical induced gap of $0.1$meV, ideally the temperature should be 10mK or lower.

\begin{figure}
\includegraphics[scale=0.38]{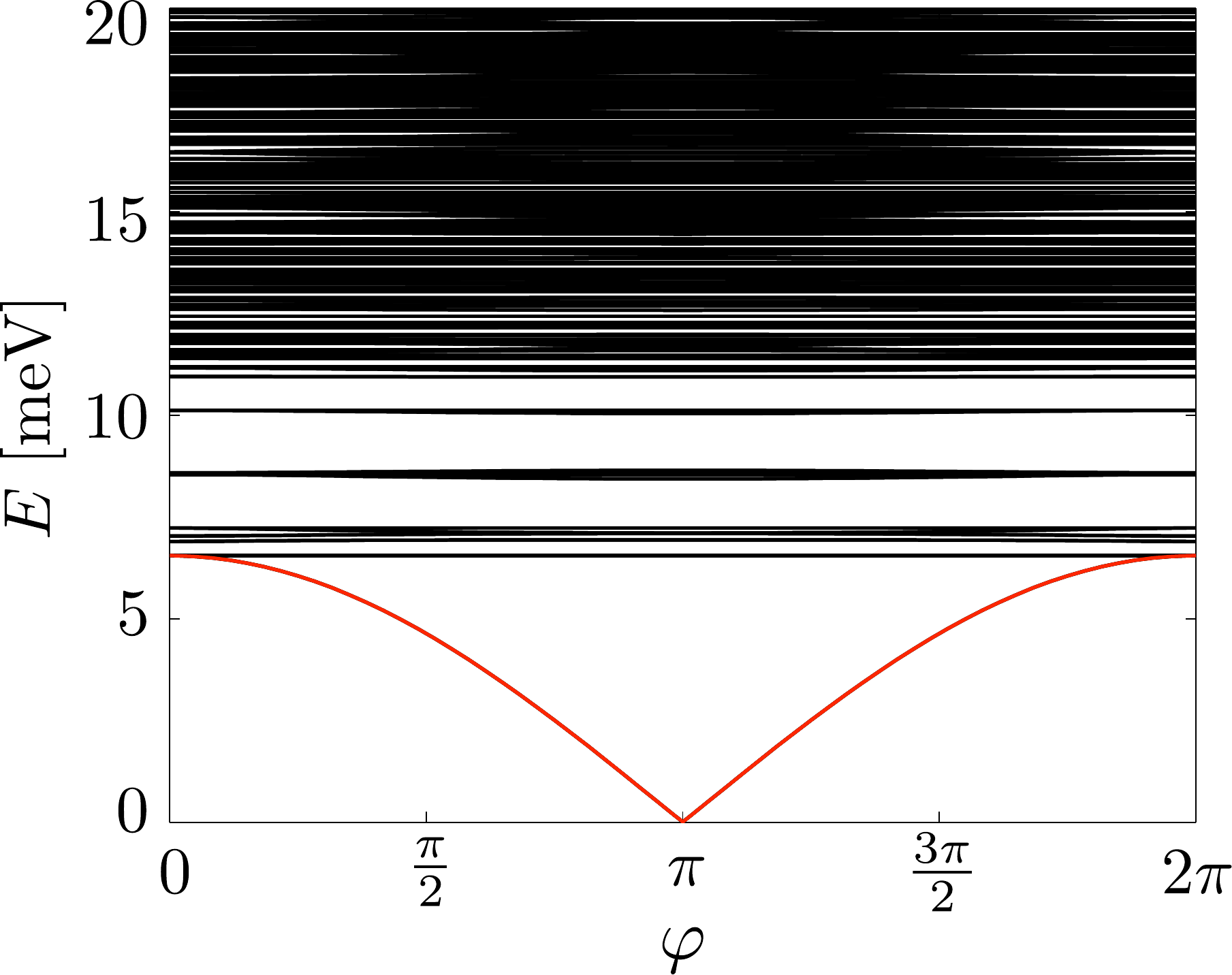} \caption{\label{fig:CPR_E} The first few quasiparticle excitation energies $E_{i}$ at $k_{z}=0$ are shown at various $\varphi$. The subgap ABS (in
red) has a kink at $\pi$ while the other states show little phase dependence. Same behaviour holds for any $k_{z}\in(-Q,Q)$;
outside of this range there is no ABS. }
\end{figure}

How do we understand this current jump? A slice of spin-up sector of the Cd$_{3}$As$_{2}$
Hamiltonian at any fixed $k_{z}\in(-Q,Q)$ combined with a slice of spin-down sector at $-k_z$ can be regarded as a 2D TI. Although a short junction on the
edge of 2D TI hosts a $4\pi$-periodic ABS with energy given by
Eq. \eqref{eq:ABS_energy}, the 2D TI embedded in Cd$_{3}$As$_{2}$ does
not locally conserve the fermion parity, so its ground state energy evolves
as 
\begin{equation}
\tilde{\varepsilon}(\varphi)=-\frac{1}{2}\Delta_{0}\left|\mbox{cos}(\frac{\varphi}{2})\right|\label{eq:kink_band}
\end{equation}
following the lower of the two particle-hole-conjugate ABS bands in order to minimize the energy. Above-gap states are not included in $\tilde{\varepsilon}(\varphi)$ because they are phase-independent and do not contribute to the supercurrent. $\tilde{\varepsilon}(\varphi)$ has a kink at $\pi$, resulting in a characteristic supercurrent jump of magnitude
\begin{equation}
\delta J=\frac{e}{\hbar}\Delta_{0} \label{eq:unit_jump}
\end{equation}
The current jump in Fig. \ref{fig:CPR} can thus be understood as the cumulative effect of many Josephson junctions mediated by the edge states of 2D TI. In Fig. \ref{fig:CPR_E}, we confirm that a slice of our model at $k_{z}=0$ indeed shows the same phase dependence as described.

With $N$ ABSs, the total current jump is
\begin{equation}
\delta J_{\text{total}}=N\delta J \label{eq:total_jump}
\end{equation}
For our numerics, $N\sim 13$ and $\Delta_{0}=10$meV, so $\delta J_{\text{total}}\sim31\mu\text{A}$, on the same order of magnitude as what Fig. \ref{fig:CPR} suggests. It is an overestimation because the actual superconducting gap as seen in Fig. \ref{fig:wide}(b) is smaller than 10meV and shrinking as $k_{z}$ approaches the nodes. This happens because the opening of surface superconducting gap is interfered by the Dirac cones in the bulk, which is inevitable when we use an unrealistically large $\Delta_{0}$ (in order to overcome the finite-size effect). In a realistic setting where $\Delta_{0}$ is much smaller than the normal-state bulk gap, we expect the superconducting gap to be uniform except in the vicinity of the nodes. In that case Eq. \eqref{eq:total_jump} provides an accurate estimate of the jump size. With realistic parameters, $N\sim 100$ , so $\delta J_{\text{total}}\sim 2.4\mu$A assuming a typical induced gap of $\Delta_0=0.1$meV. Current jump of this magnitude is readily observable in experiments. 

Due to Eq. \eqref{eq:angle}, the jump size varies as $\text{cos}(\theta)$ where $\theta$ is the angle between the junction width and the direction in which the Dirac nodes are separated ($z$-axis). To observe this, one can prepare multiple samples with different junction orientations and compare their CPR measurements.

\subsection{Non-magnetic disorder}
Disorder breaks the translational symmetry, allowing MZMs at different $k_z$'s to couple. We show that non-magnetic disorder broadens the Majorana flat bands to cover a range of near-zero energies. Consequently the discontinuity in Josephson current is rounded off but remains a robust signature for weak disorder.

Having in mind an experiment in which the chemical potential is tuned to the neutrality point, we consider only the subgap states, consisting of ABSs at the junction and bulk states on the Dirac cones. There are many ABSs; whereas the bulk density of states is vanishingly small at low energy: $D(\omega) \sim \omega^2$. Moreover, the bulk wavefunctions spread out over the entire bulk, so their probability densities near the surface are small. Based on the above arguments, we assume that any scattering between ABSs and bulk states is negligible in comparison to scattering among the ABSs, and neglect the presence of bulk states altogether.

At each $k_z\in(-Q,Q)$, there is a pair of particle-hole symmetric Andreev levels dispersing according to $E=\pm \Delta_{0} \cos \varphi/2$.
Non-magnetic disorder randomly couples $N$ pairs of ABSs such that the effective Hamiltonian reads:
\begin{align}
H = \begin{pmatrix}
I \Delta_0 \cos \varphi/2 & M \\
M^\dag & - I \Delta_0 \cos \varphi/2
\end{pmatrix}.
\end{align}
Here $M$ is a $N\times N$ random phase-independent matrix coupling the ABSs at different $k_z$'s, and $I$ is the $N\times N$ identity matrix. Note that we have chosen the shape of $H$ such that the disorder matrix resides in the off-diagonal blocks. One can show that the Hamiltonian matrix can be chosen this way by diagonalizing whichever part of the disorder matrix is in the diagonal part of the Hamiltonian and noting that the disorder potential is time-reversal invariant and cannot shift the phase difference at which the states cross zero from $\pi$. Let us also use time-reversal and particle-hole symmetries to further constrain $M$, one of the choices being $M=M^*$ and $M = - M^T$. Thus $M$ is real antisymmetric matrix and its eigenvalues are purely imaginary. The statistics of energy levels of such a random matrix are well known \cite{Beenakker1997},
\begin{align}
P(m_i) = \frac{1}{\pi c^2}\mathrm{Re\;}\sqrt{c^2 - m_i^2},\label{eq:Pmi}
\end{align}
where $c$ is a constant describing the disorder strength. We can now diagonalize $M$ and $M^\dag$ to obtain the eigenvalues of the whole matrix $H$
\begin{align}
e_i = \sqrt{\Delta_0^2 \cos^2(\varphi/2) + m_i^2}.\label{eq:ei}
\end{align}
By change of variables in the probability distribution \eqref{eq:Pmi}, we obtain disorder-averaged density of states as a function of phase difference and disorder strength:
\begin{align}
\nu(E) \propto |E| \mathrm{\; Re\;} \sqrt{\frac{c^2}{E^2 - \Delta_0^2 \cos^2 (\varphi/2)} - 1},
\end{align}

\begin{figure}
\centering{}\includegraphics[scale=0.24]{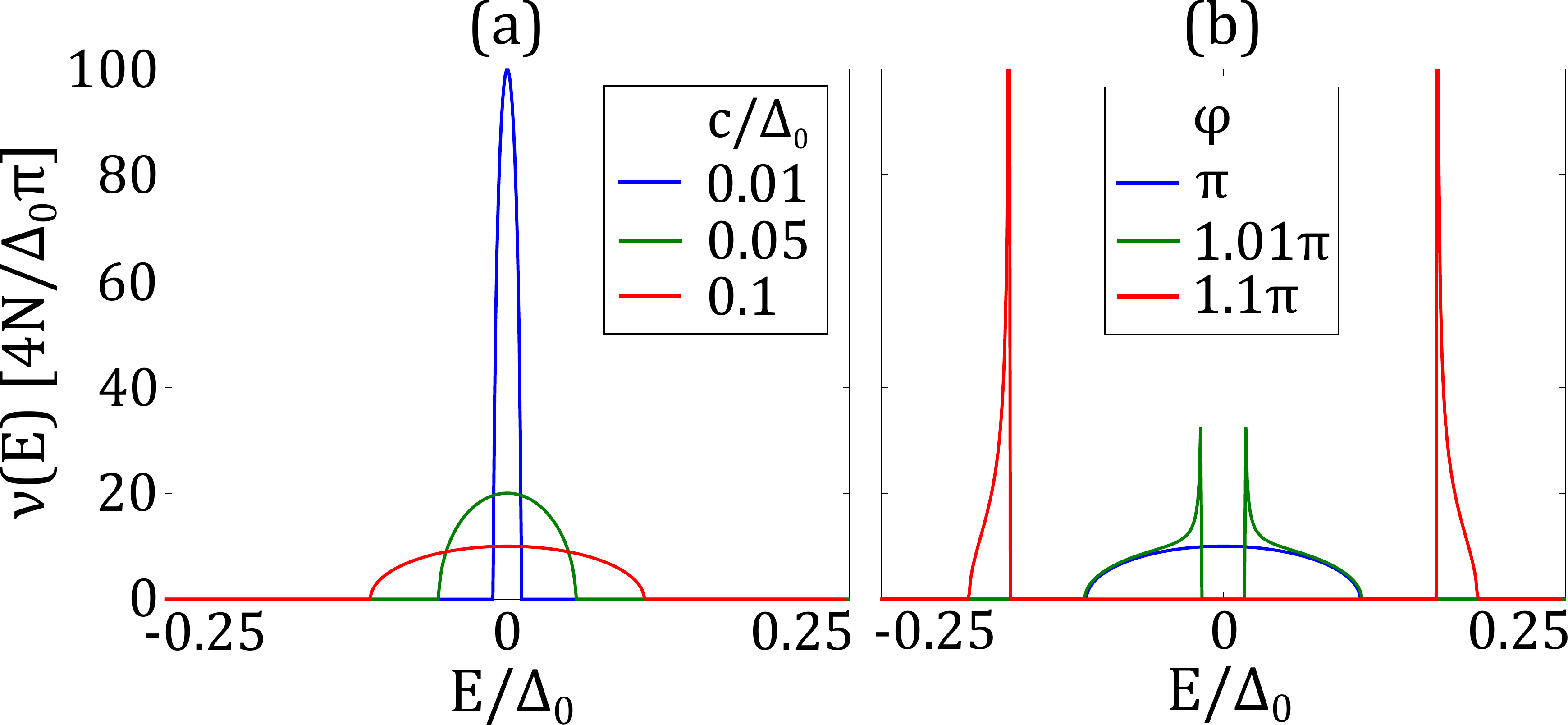}\caption{\label{fig:eq17} Panel (a) plots $\nu(E)$ at $\pi$ and shows that non-magnetic
disorder broadens the Majorana flat bands by $c/\Delta_0$. Panel
(b) shows $\nu(E)$ at $c/\Delta_{0}=0.1$
and different $\varphi$'s. The broadening diminishes as $\varphi$ deviates from $\pi$.}
\end{figure}
As shown by Fig. \ref{fig:eq17}, $\nu(E)$ describes the broadening of Andreev levels, which diminishes as $\varphi$ deviates from $\pi$, so we expect the CPR to be perturbed mostly near the jump. To confirm this, we compute the Josephson current using the free energy integrated with $\nu(E)$.
\begin{figure}
\centering{}\includegraphics[scale=0.23]{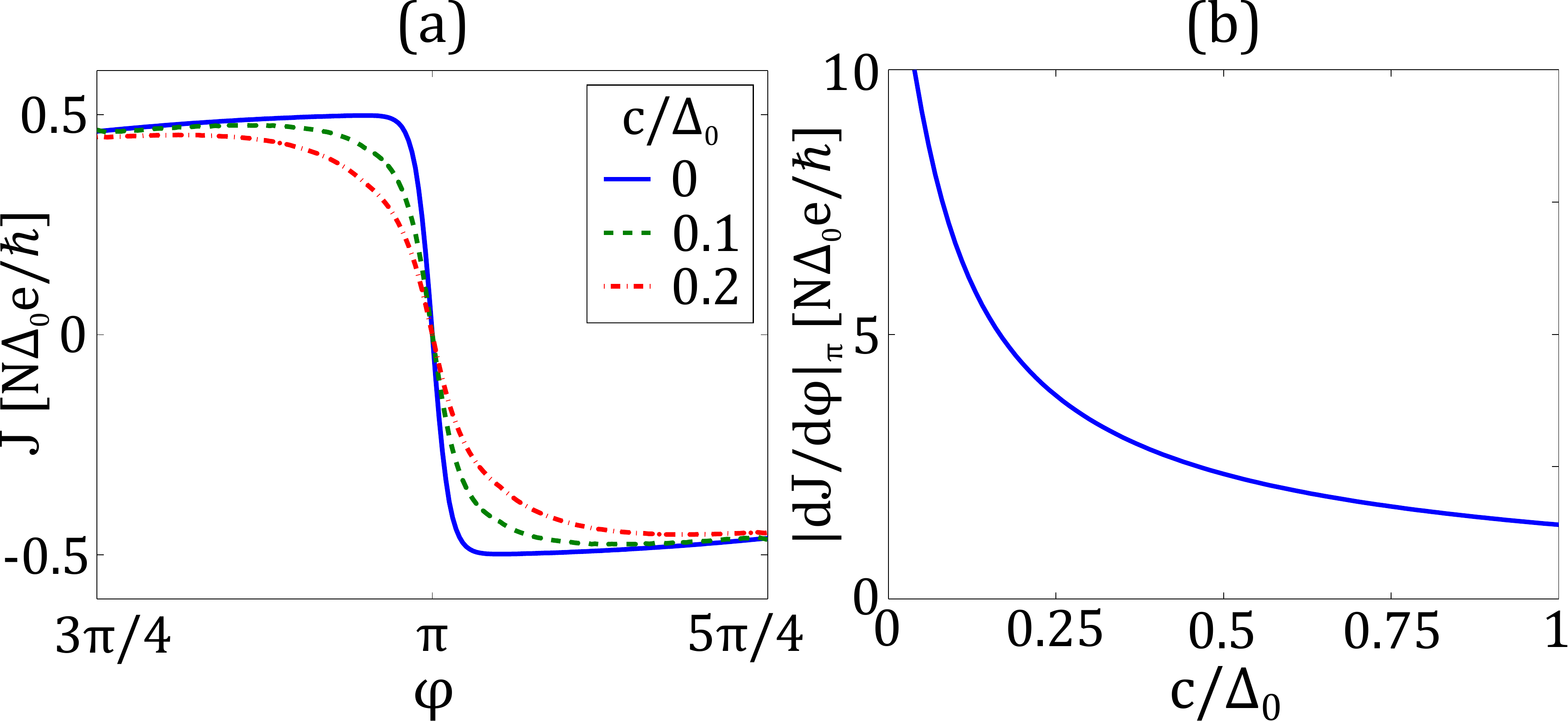}\caption{\label{fig:disorderCPR}Panel (a) shows the disordered CPR near $\pi$ at $k_{B}T/\Delta_{0}=0.01$ and various $c/\Delta_{0}$. The discontinuity is
rounded off but the jump size is unaffected. Panel (b) shows the CPR
slope at $\pi$ as $c/\Delta_{0}$ increases (again at $k_{B}T/\Delta_{0}=0.01$). For comparison, the slope at $\pi$
of a conventional sinusoidal CPR is 1.}
\end{figure}
Fig. \ref{fig:disorderCPR}(a) shows that the discontinuity is rounded off but the overall jump size is not affected. The rounding is akin to that due to nonzero temperature as seen in Fig. \ref{fig:CPR}, but to a lesser degree because $c/\Delta_0=0.1$ smoothens the discontinuity while $k_{B}T/\Delta_0=0.1$ almost completely destroys the skewness. We also measure the degree of rounding by the slope of CPR at $\pi$. It decreases with increasing $c/\Delta_0$, as shown in Fig. \ref{fig:disorderCPR}(b), but remains significantly larger than 1, the slope at $\pi$ of a conventional sinusoidal CPR.

The above RMT analysis does not consider non-averaged fluctuations, which can in principle obscure the jump. This is however not the case here. Supercurrent in the form of Eq. \eqref{eq:finite_temp} is essentially a sum of $N$ random variables when the energy eigenvalues are randomized as in Eq. \eqref{eq:ei}. By the Central Limit Theorem, the sum grows as $N$, but the standard deviation of the sum grows as $\sqrt{N}$.

\section{Conclusion}
We have shown that flat bands of MZMs occur in a linear Josephson junction
mediated by the surface states of Cd$_{3}$As$_{2}$. Numerical simulation of CPR shows a significant supercurrent jump at $\pi$, the size of which is determined by Eqs. \eqref{eq:angle}, \eqref{eq:unit_jump} and \eqref{eq:total_jump}. Low temperature ($\lesssim10$mK) and weak non-magnetic disorder mildly round off the jump.

This result applies to every $\mathcal{T}$-invariant Weyl/Dirac semimetal, which in general has an even number of Weyl nodes. With a junction of $\pi$ phase difference on the surface, each pair of Weyl nodes that used to terminate the same Fermi arc become the end points of a Majorana flat band, living in the 1D Brillouin zone parallel to the junction width. As before the CPR would have a jump, but the dependence of jump size on  junction orientation varies. For example in TaAs, a $\mathcal{T}$-invariant WSM with 24 nodes \cite{Xu2015,Lv2015}, the jump size is the largest when the junction width is parallel to the [110]-direction onto which every pair of Weyl nodes projects to two distinct points.

Tunneling spectroscopy is another way to observe the Majorana flat bands. A tunnel probe weakly coupled to the junction region is expected to measure a strong zero-bias peak in the differential conductance. Such experiments have been done to probe the ABSs of conventional junctions \cite{LeSueur2008,Pillet2010}. One might also be interested
in measuring the Fraunhofer diffraction pattern, which has been proposed to
detect the fractional Josephson effect because the $4\pi$-periodicity
of supercurrent unconventionally lifts the zeros at odd-flux quanta \cite{Lee2012}. However, in our case
the supercurrent is $2\pi$-periodic, so the Fraunhofer
pattern appears conventional, as explained in the Appendix.

\begin{acknowledgments}
The authors are indebted to NSERC and CIfAR for support. In addition
A.C. acknowledges support by the 2016 Boulder Summer School for Condensed Matter and Materials Physics through NSF grant DMR-13001648.
\end{acknowledgments}

\appendix*

\section{Fraunhofer Pattern}
An external
magnetic flux into the top surface of the Cd$_3$As$_2$ slab gives the short junction a $z$-dependent phase difference $\varphi(z)=2\pi nz/W+\varphi_{0}$,
where $n=\Phi/\Phi_{0}$ is the number of flux quanta over the junction,
$W$ is the width of the junction, and $\varphi_{0}$ is the
phase difference at $z=0$. When a constant current is driven through the junction, the junction self-tunes $\varphi_{0}$ so that the total supercurrent
\begin{equation}
I=\intop_{-W/2}^{W/2}dz\ J(\varphi(z))
\end{equation}
equals the applied current. The maximum current that the junction
can accommodate is called the critical current, $I_{c}$. In other
words, $I_{c}$ can be found by maximizing $I$ with respect to $\varphi_{0}$.
The Fraunhofer diffraction pattern refers to the single-slit-like
diffraction pattern shown by $I_{c}$ as $n$ varies. By changing
the variable, we can rewrite the integral with respect to $\varphi$:
\begin{equation}
I=\frac{W}{2\pi n}\intop_{-\pi n+\varphi_{0}}^{\pi n+\varphi_{0}}d\varphi\ J(\varphi)
\end{equation}
so the Fraunhofer pattern can be extracted from a given CPR. The limits
of the integral imply that a conventional $2\pi$-periodic CPR leads
to vanishing supercurrent at integer flux quanta. The Fraunhofer pattern is thus useful for detecting the $4\pi$-periodicity of a junction mediated by TI, dubbed the Fu-Kane junction, because the supercurrent vanishes only
at even flux quanta \cite{Lee2012}.

\begin{figure}
\centering{}\includegraphics[scale=0.4]{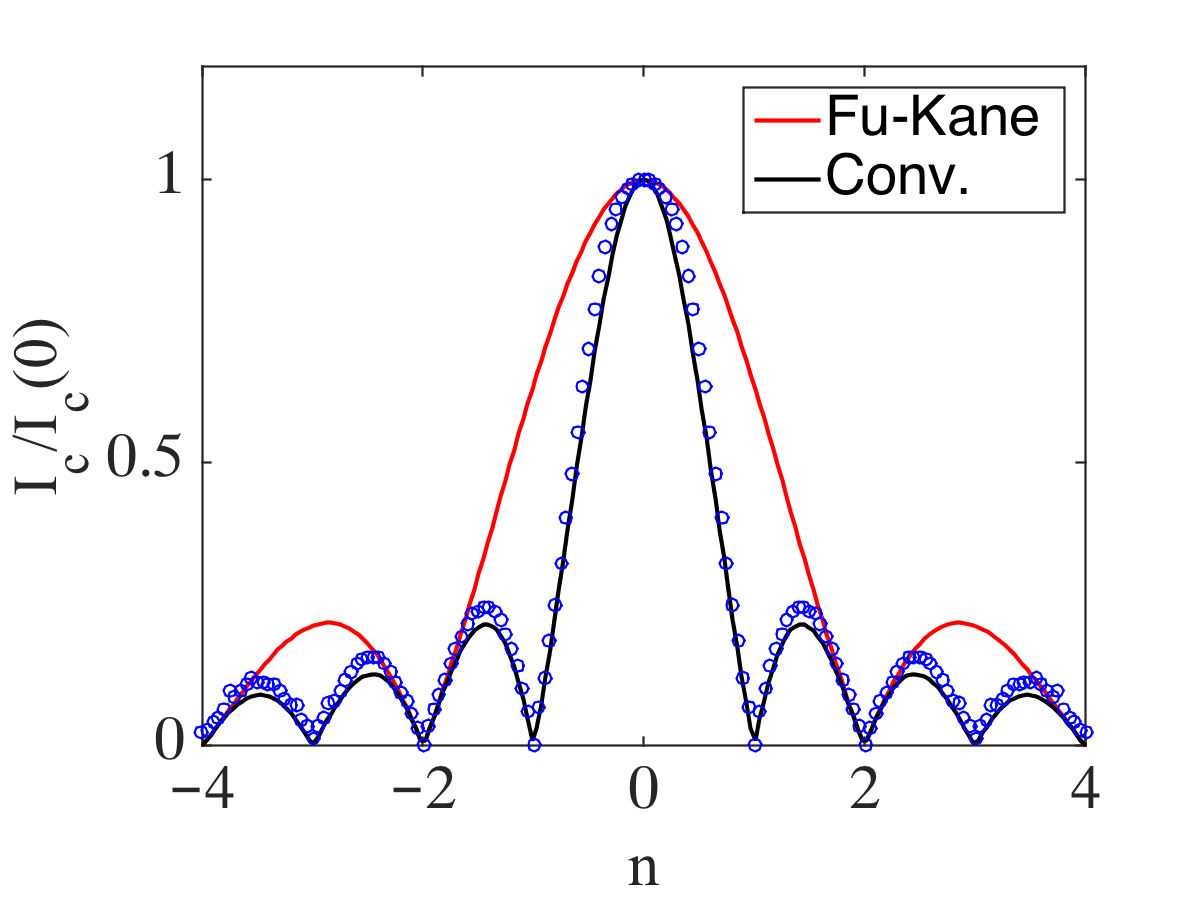}\caption{\label{fig:fraunhofer}The Fraunhofer pattern of a short junction mediated by the surface states of Cd$_{3}$As$_{2}$ is computed from the CPR in Fig. \ref{fig:CPR} and shown here by blue circles. The critical current vanishes at integer flux
quanta $n$. For comparison, we also show the Fraunhofer pattern
of the Fu-Kane junction and that of a conventional junction. In each case the critical current is normalized by its maximum.}
\end{figure}

The CPR in Fig. \ref{fig:CPR} leads to the Fraunhofer pattern in Fig. \ref{fig:fraunhofer}. It vanishes at every integer $n$ and matches the Fraunhofer pattern of a conventional junction.

\bibliographystyle{apsrev4-1}
\bibliography{My_Collection.bib}

\end{document}